\documentclass[aps,prl,twocolumn,groupedaddress]{revtex4}

\usepackage{graphicx}
\usepackage{bm}

\begin{document}
 

\title{Spatial solitons and modulational instability in the presence of large birefringence: \\ the case of highly non-local liquid crystals}


\author{Claudio Conti}
\affiliation{
Centro studi e ricerche ``Enrico Fermi,'' Via Panisperna 89/A, 00184, Rome, Italy \\and
Research center SOFT-INFM-CNR University ``La
Sapienza,'' P. A. Moro 2, 00185, Rome, Italy}
\email{claudio.conti@phys.uniroma1.it}
\author{Marco Peccianti, Gaetano Assanto}
\affiliation{
NooEL - Nonlinear Optics and OptoElectronics Laboratory, 
National Institute for the Physics of Matter, INFM-CNISM - University ``Roma Tre''
Via della Vasca Navale 84, 00146 Rome - Italy}
\homepage{http://optow.ele.uniroma3.it}
\date{\today}
\begin{abstract}
With reference to spatially non-local nematic liquid crystals, we develop a theory of optical spatial solitons and modulational instability in anisotropic media with arbitrarily large birefringence.
Asymmetric spatial profiles and multivalued features are predicted for self-localized light versus walk-off angle. 
The results hold valid for generic self-focusing birefringent media and
apply to large angle steering of individual and multiple self-trapped optical beams.
\end{abstract}
\maketitle
An appealing approach towards the
realization of digital multidimensional all-optical processors and information routers is the use of spatially self-trapped optical filaments -or solitons- as readdressable light pencils able to guide signals in arbitrary directions. Although recent years have witnessed widespread investigations 
of spatial solitons in various nonlinear systems
\cite{Trillo01,KivsharBook}, to date a proper description and, hence, modeling and prediction of such functionality is hampered by two main issues.
The first is somewhat of a technical origin: while the most studied nonlinearities 
for stable two-dimensional spatial solitons rely on birefringent materials, \cite{StegemanSegev}
the effects of anisotropy have been accounted only for small walk-off or propagation along one of the principal axes
 (e.g. in photorefractives
or quadratic media). \cite{Zozulya98,Krolikowski98, Polyakov2002,Torner98, Rosanov2001,Guo2000} 
Dealing with anisotropy in a perturbative way,
current models fail to predict the formation of 
self-collimated beams readdressable over wide-angles. The second issue relates to the ubiquitous paraxial 
approximation, which is unable to treat propagation at large angles with respect to the 
input wavevector, i.e. to the launch direction $\hat{\bf z}$ of the beam 
generating the soliton. 

The above considerations hold valid also for spatial modulational instability (MI), a process tipically accompanying (or precurring) solitons. MI describes
 unstable plane-waves which, through self-focusing, break up into transversely periodic patterns eventually evolving into filaments. \cite{Peccianti03} 
To date, the analysis of optical MI has been limited to small birefringence and paraxial behavior even in crystals with significant anisotropy. 

In this Letter we address nematic liquid crystals (NLC) as a natural environment to assess the role of a strong anisotropy
in beam self-localization and MI. 
In doing so, by letting 
the beam be paraxial not with respect to  $\hat{\bf z}$ but to a rotated reference system, we develop a model which properly accounts for arbitrarily large 
walk-off and birefringence. While this allows to deal with beam steering over large angles, \cite{PecciantiNature} it also enlightens unexplored
features of solitons and MI for arbitrary crystal orientations. The beam,
polarized as an extraordinary wave, gives up radial symmetry while acquiring an asymmetric transverse profile. The latter, depending on angle of propagation, reveals a nontrivial distribution of both longitudinal and transverse 
components.  Moreover, differing anisotropic solitons or MI patterns can be expected for a given walk-off.

The approach we introduce hereby is general and can be applied to any anisotropic nonlinear medium. For illustration sake and in order to pin-point a physically relevant system, we explicitly refer to a voltage-biased glass cell containing a thick layer of planarly-anchored nematic liquid crystals.  Such configuration, encompassing a significant and externally-adjustable degree of birefringence (and walk-off), has been previously exploited for various experiments with spatial solitons ({\it nematicons}) and MI. \cite{Peccianti03, PecciantiNature, OPN, Conti04}  In the present context we adopt a reference system and notation as in Fig. 1.
The starting point is the vectorial wave equation:
$\nabla\times\nabla\times{\mathbf E}=k_0^2\,{\bm \varepsilon}\cdot{\mathbf E}$, with (constant) dielectric tensor ${\bm \varepsilon}$.
Looking for a plane-wave solution ${\bf E}={\bf A}\exp(i k_0 n z)$ 
propagating along $z$, 
the linear homogeneous algebraic system 
\begin{equation}
\label{Ldefinition}
\mathcal{L}(n)\cdot{\bf A}=[n^2 (\hat{\bf z}\hat{\bf z}-{\bf I})+{\bm \varepsilon}]\cdot {\bf A}=0\text{,}
\end{equation}
where ${\bf I}$ is the identity matrix, $\hat{\bf z}$ the unit vector in the $z-$direction, 
and $\hat{\bf z}\hat{\bf z}$ the dyadic tensor whose elements are $\hat{z}_i \hat{z}_j$ with with $i,j=\{x,y,z\}$,  yields the allowed values for $n$, i.e. 
ordinary and extraordinary refractive indices.
\begin{figure}
\includegraphics[width=7.0cm]{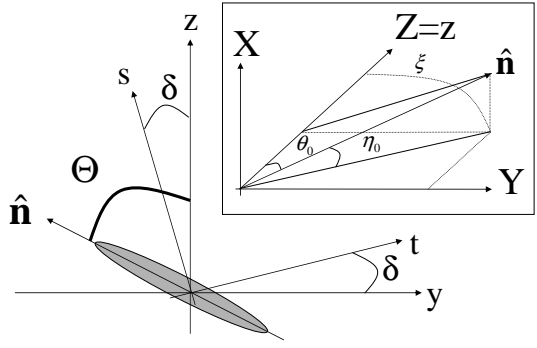}
\caption{
Adopted coordinate system: The grey ellipse is a sketch of a LC molecule, $\hat{\bf n}$ is its director, $s$ and $t$ are the walk-off direction and its normal, respectively. $\Theta=\theta_0$ in the absence of optical excitation,
$\delta(\theta_0)$ is the walk-off angle. 
The inset illustrates a feasible experimental arrangement, with $X$, $Y$ and $Z=z$ the axes in the laboratory frame.
The applied voltage determines the elevation angle $\eta_0$ of the molecules, which at zero bias lie in the $Y,Z$ plane with azimuth $\xi$. The beam propagates along $\hat{\bf s}$ in the plane ($\hat{\bf n}$,$\hat{\bf z}$).
\label{figuredirector}}
\end{figure}
Considering a light beam propagating in the midplane of a much thicker cell, for NLC with director $\hat{\bf n}$ (e.g. mean orientation of molecular major axes) in the $(y,z)$ plane, as in figure \ref{figuredirector}, 
the relative permittivity tensor is given by $\varepsilon_{ij}=\varepsilon_\perp \delta_{ij}+\Delta\varepsilon n_i n_j$,  and the anisotropy $\Delta\varepsilon$ constant in the illuminated region.
The ordinary (o-) wave is polarized along $x$ and $n_o^2=\varepsilon_\perp$, while the extraordinary (e-) wave belongs to the $(y,z)$ plane
and the resulting index is 
$n_e(\theta_0)^2=2\varepsilon_\perp(\varepsilon_\perp+\Delta\varepsilon)/[2\varepsilon_\perp+\Delta\varepsilon+\Delta\varepsilon cos(2\theta_0)]$.
The unit-vector associated to the e-wave is denoted $\hat{\bf t}(\theta_0)$, and its normal defines 
the walk-off direction $\hat{\bf s}(\theta_0)$. $\delta(\theta_0)$ is the walk-off angle, with
$\tan(\delta)=[\Delta\varepsilon \sin(\theta_0)\cos(\theta_0)]/[\varepsilon_\perp+ \Delta\varepsilon \cos(\theta_0)^2]$.
We omit hereafter the dependence on $\theta_0$.

The general plane-wave solution with wavevector parallel to $z$ is a combination of e- and o-waves. 
In the following, we only consider the e-wave 
polarized along $\hat{\bf t}$ and propagating along $\hat{\bf s}$; the ordinary wave, being orthogonal to $\hat{\bf n}$, at the lowest order of approximation
does not affect NLC molecular re-orientation through dipole-field interaction, because of the existence of a threshold known as the optical Freedericksz transition. \cite{Tabirian86,Khoo95}
Hence, for o-waves of intensity well below the Freedericksz threshold, the e-wave is the leading term in the NLC reorientational nonlinear response.

\begin{figure}
\includegraphics[width=5.0cm]{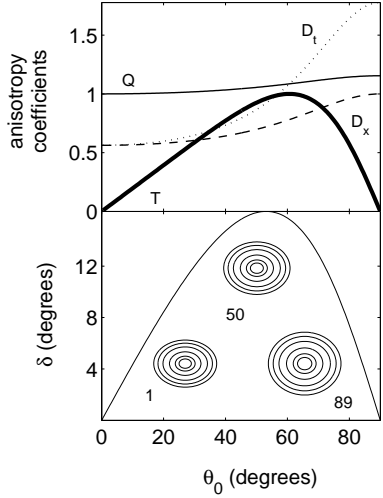}
\caption{(Top) Dimensionless anisotropy coefficients Vs $\theta_0$, see text.
(Bottom) Walk-off angle $\delta$ Vs $\theta_0$; the insets show the soliton transverse intensity profiles for various $\theta_0$ (labeled in degrees).
Parameters: $\varepsilon_\perp=(n_0)^2=1.5^2$; $\Delta\varepsilon=1.75$ ($max(n_e)=n_0+0.5$). \label{figureanisotropy}}
\end{figure}

The optical field perturbs the dielectric tensor as 
$\bm{\varepsilon}\rightarrow\bm{\varepsilon}+\epsilon^2\bm{\delta\varepsilon}$, with $\epsilon$ a {\it smallness 
parameter} to be taken equal to $1$ at the end of the derivation.
The expansion is written as 
${\bf E}=\left[ \hat{\bf t} E_e+
\epsilon {\bf F}_e+\epsilon^2 {\bf G}_e+...\right]\exp(i k_0 n_e z)$,
with $E_e$, ${\bf F}_e$ and ${\bf G}_e$ depending on multiple slow
scales 
$x_n=\epsilon^n x$, $t_n=\epsilon^n t$ and $s_n=\epsilon^n s$ ($n=1,2,...$)
in the reference system $(x,t,s)$.
At the order $O(\epsilon)$, it is 
\begin{equation}
\label{eq2}
\begin{array}{l}
k_0^2 {\bf\mathcal{L}}(n_e)\cdot{\bf F}_e=
i k_0 n_e [\hat{\bf z}\times\nabla_1\times(E_e \hat{\bf t})+\nabla_1\times\hat{\bf z}\times(E_e \hat{\bf t})]=\\
i k_0 n_e \hat{\bf t} [-2 \cos(\delta) \frac{\partial E_e}{\partial s_1}]+i k_0 n_e \hat{\bf x} [\sin(\delta) \frac{\partial E_e}{\partial x_1}]+\\
i k_0 n_e \hat{\bf s} [\sin(\delta) \frac{\partial E_e}{\partial s_1}+\cos(\delta) \frac{\partial E_e}{\partial t_1}]\text{.}
\end{array}
\end{equation}
The solvability condition implies the rhs of (\ref{eq2}) to be orthogonal to the null space of $\mathcal{L}(n_e)$, given by $\hat{\bf t}$:
$\partial E_e /\partial s_1=0$. For the first-order vectorial correction ${\bf F}$, writing $\mathcal{L}(n_e)$ in the $(x,t,s)$ system provides:
$F_e^x=(i n_e \sin \delta/k_0 \lambda_x) \partial E_e/ \partial x_1$, $F_e^t=0$,
$F_e^s=( i n_e \cos \delta/k_0 \lambda_s)  \partial E_e/\partial t_1$;
being $\lambda_{x,s}$ the non-vanishing eigenvalues of $\mathcal{L}(n_e)$ (see Eqs. (\ref{Dcoeffs}) below).
Hence, at this order of approximation, the electric field \textit{is not linearly polarized as an extraordinary wave}, but its polarization varies 
across the finite trasverse profile.
At the order $O(\epsilon^2)$
\begin{equation}
\begin{array}{l}
k_0^2{\bf\mathcal{L}}(n_e)\cdot{\bf G}_e=-k_o^2 \bm{\delta\varepsilon}\cdot\hat{\bf t}E_e+  \nabla_1\times\nabla_1\times(E_e \hat{\bf t})\\
i k_0 n_e [\hat{\bf z}\times\nabla_1\times{\bf F_e}+\nabla_1\times\hat{\bf z}\times{\bf F_e}]+\\
i k_0 n_e [\hat{\bf z}\times\nabla_2\times(E_e \hat{\bf t})+\nabla_2\times\hat{\bf z}\times(E_e \hat{\bf t})]
\text{.}
\end{array}
\end{equation}
Using the result obtained at the previous order, from the solvability condition
$\hat{\bf t}\cdot{\bf\mathcal{L}}(n_e)\cdot{\bf G}_e=0$
it is found (in the original scales, $\epsilon\rightarrow1$)
\begin{equation}
\label{ewaveMMS}
2 i k_0 n_e cos(\delta) \frac{\partial E_e}{\partial s}+D_{t}\frac{\partial^2 E_e}{\partial t^2}+
D_{x}\frac{\partial^2 E_e}{\partial x^2}+k_0^2 (\hat{\bf t}\cdot\bm{\delta\varepsilon}\cdot\hat{\bf t}) E_e=0
\end{equation}
i.e., the paraxial propagation equation in the walk-off system. The modified diffraction coefficients are 
\begin{equation}
\label{Dcoeffs}
\begin{array}{l}
D_{t}=\displaystyle\frac{n_e^2\cos(\delta)^2}{\lambda_s}=\\
\frac{\varepsilon_\perp (\Delta\varepsilon+\varepsilon_\perp)[\Delta\varepsilon+2\varepsilon_\perp+\Delta\varepsilon \cos(2\theta_0)]^2}
{[\Delta\varepsilon^2+2\Delta\varepsilon\,\varepsilon_\perp+2\varepsilon_\perp^2+\Delta\varepsilon
(\Delta\varepsilon\varepsilon_\perp)\cos(2\theta_0)]^2}\\
D_{x}=\displaystyle\frac{n_e^2\sin(\delta)^2}{\lambda_x}=\\
\frac{\varepsilon_\perp[\Delta\varepsilon+2\varepsilon_\perp+\Delta\varepsilon\,cos(2\theta_0)]}{
\Delta\varepsilon^2+2\Delta\varepsilon\varepsilon_\perp+2\varepsilon_\perp^2+\Delta\varepsilon(\Delta\varepsilon+2\varepsilon_\perp)cos(2\theta_0)}
\text{.}
\end{array}
\end{equation}
$D_{t}\neq D_{x}$ involves the absence of radially symmetric spatial solitons, 
with ellipticity (ratio between waists across $t$ and $x$, respectively) given by $Q\equiv(D_{t}/D_{x})^{1/4}$ (see below). 
Noteworthy, when the birefringence $\Delta\varepsilon\rightarrow0$, it
is $D_{x}=D_{t}=1$ and isotropic propagation is retrieved.
Figure \ref{figureanisotropy} plots these quantities versus $\theta_0$ for a highly-birefringent NLC.\cite{Gauza2003}
We need to stress that paraxiality in the walk-off system does not imply paraxiality in the original reference
$(x,y,z)$, as witnessed by the fact that, when re-writing equation (\ref{ewaveMMS}) in $(x,y,z)$ the second-order derivatives with respect to $z$ re-appear.
Since (\ref{ewaveMMS}) holds for any walk-off, this treatment can model wide angle steering of spatial solitons, as e.g. obtainable by exploiting the voltage dependence of $\delta$ in the NLC geometry.

The molecular director $\hat{\bf n}$ lies in the $(y,z)$ plane and can be expressed in terms of angle $\Theta=\theta_0+\Psi$ (see figure \ref{figuredirector}).
Using a multiple scale expansion, 
for the e-wave perturbation along $\hat{t}$
at the lowest order in $\Psi$ we have 
$\delta\varepsilon_{tt}\equiv \hat{\bf t}\cdot\bm{\delta\varepsilon}\cdot\hat{\bf t}=\Delta\varepsilon\,T(\theta_0)\Psi$
with 
(see figure \ref{figureanisotropy})

\begin{equation}
\!\!T=\frac{2\varepsilon_\perp(\Delta\varepsilon+\varepsilon_\perp)\sin(2\theta_0)}
{(\Delta\varepsilon+\varepsilon_\perp)^2+\varepsilon_{\perp}^2+
[(\Delta\varepsilon+\varepsilon_\perp)^2-\varepsilon_{\perp}^2]\cos(2\theta_0)}\text{.}
\end{equation}
The NLC orientation is described by the functional $F=F_K+F_{RF}+F_{opt}$,
with $F_K$ (Frank-term) accounting for the elastic properties of the NLC and $F_{RF}$ for its director distribution (and hence $\theta_0$) due to an externally applied (electric or magnetic) field in the absence of 
light. \cite{Khoo95,SimoniBook}
For a dominant e-wave, the optical contribution to the energy $F$ is 
$F_{opt}=-[\varepsilon_\perp |E_e|^2+\Delta\varepsilon(\hat{\bf  n }\cdot\hat{\bf t})^2 |E_e|^2]/4$.

In the single-constant approximation (i.e., $K=K_1=K_2=K_3$ for molecular splay, bend and twist, respectively), from the 
Fr\'echet derivative of $F$ and at the lowest-order in $\Psi$  we get:\cite{SimoniBook,Khoo95,Conti03}
\begin{equation}
\label{angles}
K\nabla^2 \Psi
-A(\theta_0) \Psi
+\frac{\epsilon_0\Delta\epsilon}{4}\sin[2(\theta_0-\delta)]|E_e|^2=0\text{.}
\end{equation}
$A(\theta_0)$ is determined by cell geometry and bias.
For the configuration in figure \ref{figuredirector}, $\theta_0$ is determined by the voltage-driven elevation $\eta_0$ in the middle of the cell and by the azimuth $\xi$ due to NLC anchoring (at the interfaces defining the cell) with respect to $z$ \cite{Conti03}. We obtain 
\begin{equation}
\label{Adef}
A(\theta_0)=\frac{\varepsilon_0\Delta\varepsilon_{RF}}{cos(\xi)^2}\left(\frac{V}{L}\right)^2 
\left[\frac{\sin(2\theta_0)}{2\theta_0}-cos(2\theta_0)\right]\text{,}
\end{equation}
with $\Delta\varepsilon_{RF}$ the low-frequency (relative) permittivity, $L$ an effective cell thickness
over which the voltage $V$ is applied and $\cos(\theta_0)=\cos(\xi)\cos(\eta_0)$. 

\begin{figure}
\includegraphics[width=8.0cm,angle=0]{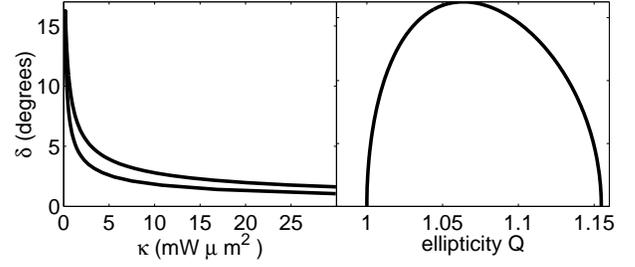}
\caption{(Left) $\delta$ Vs $\kappa$ and (Right) $\delta$ Vs ellipticity, showing the  soliton families 
attainable for the same walk-off. In the adopted units $\kappa$ measures the power in $mW$ 
needed for a $1\mu m$ waist soliton. \label{figurefeatures}}
\end{figure}

\noindent {\it Spatial solitons.}
Optical spatial solitons or \textit{nematicons} \cite{OPN} are solutions of equations (\ref{ewaveMMS}) and (\ref{angles})
in the form $E_e=(2Z_0/n_e)^{1/2}U(x,t)\exp(i\beta s)$, with $U^2$ the intensity profile 
and $\beta$ the 
"nonlinear wavevector,"  and $\partial_s\Psi=0$ (hereafter, we will always take $\partial_s\Psi=0$ in (\ref{angles}), since the optical field is slowly varying along $s$, see 
also \cite{Conti03}).
Self trapped beams travel along $\hat{\bf s}$, while their phase profile is orthogonal to the plane 
$k_0 n_e z+\beta s=[k_0 n_e+\beta cos(\delta)] z-\beta sin(\delta) y=$constant, implying that for small $\beta$ (low-power solitons) 
the phasefront corresponds to a plane wave propagating along $z$, gradually tilted towards $s$ as the power increases.
In other words, the nonlinearity tends to reshape the extraordinary wave into an ordinary-like configuration, by distorting
the phase-fronts towards the plane orthogonal to the Poynting vector.
The exact soliton profiles $U$ can be obtained numerically.
Nevertheless, relevant insights 
can be obtained 
in the highly non-local limit,\cite{Snyder97,Krolikowski00} as applicable to NLC.\cite{Conti03} 
For NLC as in actual experiments, in fact, 
the perturbation $\Psi$ extends far from the excitation, so that the beam essentially experiences an index perturbation 
with a parabolic-like shape.\cite{Conti04} Writing $\psi\cong\psi_0+\psi_2 (x^2+t^2)$, the equation for $U$ can be 
analytically solved by separation of variables, yielding a wide class of self-trapped solutions including higher-order and breathing ones. The simplest profile is gaussian with intensity profile:
\begin{equation}
\label{finalgaussian}
\mathcal{I}=\frac{P^2}{\pi \kappa (D_x D_t)^{1/4}}\exp[-\frac{P}{\kappa}( \frac{t^2}{\sqrt{D_{tt}}}+\frac{x^2}{\sqrt{D_{xx}}})]\text{,}
\end{equation}
where $P$ is the soliton power, and $\kappa$ is the constant of the existence curve: 
$P w_0^2=\kappa$, with $w_0$ the intensity ($1/e$) waist in
the isotropic limit ($D_x=D_t=1$). 
It is 
$\kappa=2 K n_e c \lambda^2 \sqrt{D_x D_t}/\pi \Delta\varepsilon^2 T sin(2\theta_0 -2\delta)$.
In deriving (\ref{finalgaussian}), we used $\psi_2=-\mathcal{I}_0 \Delta\varepsilon\sin[2(\theta-\delta)]/8K n_e c$, as found
from (\ref{angles}) with $\mathcal{I}_0$ the peak intensity when $A\rightarrow0$ (highly nonlocal regime).
The self-trapped beam travels at any angle $\delta(\theta_0)$ with
a gaussian profile and ellipticity $Q=(D_{t}/D_{x})^{1/4}$, as anticipated.
As in the case of MI (see below), two solitons (with different $\kappa$) exist
for the same $\delta$: their family is multivalued (one for each $\kappa$, spanned by the power $P$) 
with respect to walk-off $\delta$ (and unfolded by $\theta_0$, i.e. propagating in different planes), as visible in figure \ref{figurefeatures}.
Noticeably, a straightforward generalization of the theory in \cite{Turitsyn85} (see also \cite{Bang02}) enables us to state that such solitons 
(i.e. those of eqs. (\ref{ewaveMMS}) and (\ref{angles})) are unconditionally stable. Indeed, the Hamiltonian
for the system is written as
$H=\int D_t |\partial_t E_e|^2 +D_x |\partial_x E_e|^2 dx dt+ H_{nl}$
where $H_{nl}$ the nonlinear nonlocal part which is identical to the isotropic case, and  bounded from below as shown in \cite{Turitsyn85};
 the remaining part is also bounded because, from Eq. (\ref{Dcoeffs}),
$D_t$ and $D_x$ are not smaller than $\varepsilon_\perp/(\varepsilon_\perp+\Delta\varepsilon)$ (see e.g. figure \ref{figureanisotropy}) and this implies the boundness of $H$, and hence the stability.

\begin{figure}
  \includegraphics[width=6.5cm,angle=0]{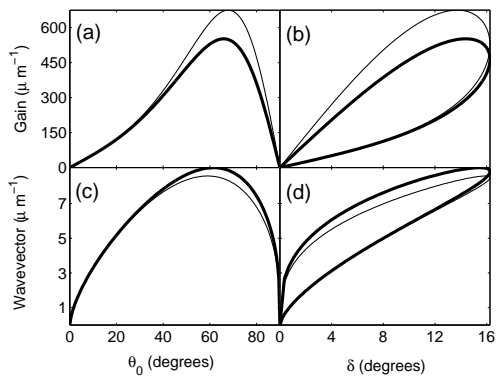}
\caption{(a) Maximum MI-gain Vs $\theta_0$; (b) as in (a) Vs walk-off $\delta$; (c) maximally amplified 
spatial harmonic Vs $\theta_0$; (d) as in (c) Vs $\delta$. Thick (thin) lines refer to $l=x$ ($l=t$).
Parameters: $V=1V$, $L=75\mu m$, $\xi=0$,
$K=10^{-11}N$, $\varepsilon_\perp=2.25$, $\Delta\varepsilon=1.75$, 
$E_0=5\times10^4 V m^{-1}$ and $A$ is given by Eq.(\ref{Adef}).\label{figureMI}}
\end{figure}

\noindent {\it One-dimensional modulational instability.}
A sufficiently-wide elliptic beam propagating along $z$ in the nonlinear sample  approximates well a one-dimensional plane-wave. 
Owing to reduced diffraction across the major axis of the ellipse, in fact, the nonlinearity acts 
mainly in one-dimension. As confirmed by experiments, \cite{Peccianti03} MI causes the initially uniform beam profile to break-up of into a periodic pattern and, eventually, into periodically-spaced filaments. 
The instability can be intuitively described as the (selective in transverse spatial frequency) 
amplification of small amplitude noise superimposed to the input beam.
For simplicity, we consider two limiting cases: an input 
ellipse with long axis oriented i) along $x$ ($l=x$) or ii) along $t$ ($l=t$). 
Plane-wave noise components of wavevector $k_{x,t}$ will grow in amplitude along $s$ with gain 
$\exp[g_l(k_l)s]$, being $l=\{x,t\}$: hence {\it filaments form along the walk-off direction}. A standard approach \cite{Krolikowski01,Peccianti03} provides:
\begin{equation}
\label{MIgain}
g_l=\frac{\sqrt{D_{l}} k_l }{2 k_0 n_e \cos(\delta)}
\sqrt{
\frac{E_0^2 k_0^2 \epsilon_0 \Delta\epsilon^2 T \sin[2(\theta_0-\delta)]} {2A+2K k_{l}^2}-D_l k_l^2}\text{,}
\end{equation}
being $E_0$ the peak amplitude of the pump (input) field.
Eq.(\ref{MIgain}) is the generalized expression of MI-gain in the presence of both non-locality and anisotropy, and reduces to the known result 
\cite{Peccianti03} in the isotropic regime.
The MI-gain is spectrally affected by both the spatial orientation of the input ellipse (with major axis parallel to either $x$ or $t$ axes in cases i) and ii), respectively) and $\theta_0$. Figure \ref{figureMI} shows the calculated gain profile for either orientations Vs $\theta_0$
(i.e. by varying the cell bias) and typical NLC parameters.
Notably, the peak gain corresponds to the maximum walk-off; when graphed Vs $\delta$, however, MI is multivalued: for 
the same $\delta$ two distinct wave-patterns can emerge, 
belonging to different planes $(t,z)$ and corresponding to different $\theta_0$. 
In an advanced stage of MI-induced filamentation, this implies the possibility of angularly steering an entire array of regularly spaced (soliton) channel waveguides by acting on $\theta_0$. 

In conclusion, by developing a comprehensive model for nonlinear wave propagation in the presence of significant walk-off, we predict the existence of (multi-value) spatial solitons and modulational instability in 
highly birefringent and non-local media. Self-trapped beams travel at arbitrarily large walf-off angles, 
which determine their elliptic intensity profile.  The results hold valid for individual solitons and arrays of filaments as generated through modulational instability, and can
be readily extended to other media. 
In NLC, where walk-off can be adjusted by an external voltage, 
wide angle bias-controlled steering of ultra-thin anisotropic solitons 
could be effectively implemented, leading to applications such as 
optical information processing in space (e.g. spatial de-multiplexing) and optical tweezers.


\end{document}